\documentclass[12pt]{article}

\usepackage{scicite}
\usepackage{graphicx}
\usepackage{wrapfig}
\usepackage[hyphens]{url}
\usepackage{times}
\usepackage{tabu}
\usepackage{longtable}
\usepackage{amsmath}

\newenvironment{sciabstract}{%
\begin{quote} \bf}
{\end{quote}}

\title{Effects of Racial Segregation on Economic Productivity in U.S. Cities}

\author
{Andrew J. Stier$^{1}$$^\ast$, Sina Sajjadi$^{2,}$$^{3}$, Lu\'is M.A. Bettencourt$^{4,}$$^{5}$,\\ Fariba Karimi$^{2}$, Marc G. Berman$^{1,}$$^{6}$$^\ast$\\ \\
\normalsize{$^{1}$Department of Psychology, University of Chicago, Chicago, IL, USA}\\ 
\normalsize{$^{2}$Complexity Science Hub, Vienna, Austria}\\ 
\normalsize{$^{3}$Central European University, Vienna, Austria}\\ 
\normalsize{$^{4}$Department of Ecology and Evolution, University of Chicago, Chicago, IL, USA}\\ 
\normalsize{$^{5}$Mansueto Institute for Urban Innovation, University of Chicago, Chicago, IL, USA}\\ 
\normalsize{$^{6}$The University of Chicago Neuroscience Institute, University of Chicago, Chicago, IL, USA}\\ \\
\normalsize{$^\ast$To whom correspondence should be addressed; E-mail:}\\
\normalsize{andrewstier@uchicago.edu or bermanm@uchicago.edu.}
}

\date{}

\begin{document} 


\baselineskip24pt


\maketitle

\begin{sciabstract}
Homophily and heterophobia, the tendency for people with similar characteristics to preferentially interact with (or avoid) each other are pervasive in human social networks. Here, we develop an extension of the mathematical theory of urban scaling which describes the effects of homophily and heterophobia on social interactions and resulting economic outputs of cities. Empirical tests of our model show that increased residential racial heterophobia and segregation in U.S. cities are associated with reduced economic outputs and that the strength of this relationship increased throughout the 2010s. Our findings provide the means for the formal incorporation of general homophilic and heterophobic effects into theories of modern urban science and suggest that racial segregation is increasingly and adversely impacting the economic performance and connectivity of urban societies in the U.S.
\end{sciabstract}


\section*{Introduction}
Homophily and heterophobia, the tendency of more similar individuals to preferentially interact and avoid interactions with others, are intimately familiar: most real-world~\cite{mcpherson2001birds} and digital~\cite{thelwall2009homophily} social networks show some degree of increased connectivity within certain groups and decreased connectivity between groups. Whether these preferences occur across characteristics of morality~\cite{dehghani2016purity}, race~\cite{mollica2003racial}, or nationality~\cite{thelwall2009homophily}, minor individual preferences amplified by structural proximity can result in large group-level  differences~\cite{kossinets2009origins,schelling1971dynamic,schelling2006micromotives,dall2008statistical}. Moreover, observed group differences in connectivity (outcome homophily/heterophobia) can have tremendous impacts on human behavior~\cite{karimi2018homophily,lee2019homophily} despite not always being the result of strong individual choice preferences. In general, these effects expose individuals to less variety, knowledge and choice so that they slow down learning~\cite{golub2012homophily}, increase cognitive biases~\cite{lee2019homophily}, and limit the spread of information~\cite{halberstam2016homophily} among other detrimental effects~\cite{ertug2022does,ibarra1992homophily}.

In cities, these effects play out across various types of preferences: people tend to travel between neighborhoods with similar socioeconomic demographics~\cite{heine2021analysis}, patterns of violent crime~\cite{graif2017neighborhood}, and overall well-being~\cite{lathia2012hidden}. There is some evidence that spatial proximity in central locations can combat homophily and heterophobia, suggesting that these effects play out over large distances and more peripheral settings in cities~\cite{xu2019quantifying}. In addition, long histories of racism have led to spatial segregation among racial groups, particularly in the United States (U.S.)~\cite{kruse2013white,nardone2020historic}. However, despite the universality of these effects in urban environments and their many pernicious effects, homophily and heterophobia have not yet been formally incorporated into the theoretical framework of modern urban science~\cite{bettencourt2013origins,bettencourt2021introduction}, which often assumes homogeneous (non-homophilic/heterophobic) mixing. Here we begin this process by developing homophily and heterophobia adjustments to the equations of urban scaling theory. We validate these adjustments empirically and provide evidence that racial heterophobia at the city level is predictive of lower overall economic productivity in the U.S. and that the strength of this relationship increased throughout the 2010's.

\section*{Results}
\subsection*{Homophily and Heterophobia in Urban Scaling Theory}
Urban scaling theory~\cite{bettencourt2013origins,bettencourt2021introduction}, which provides a theoretical backbone of modern urban science, describes cities as spatially embedded networks of socioeconomic interactions. Cities arise when the benefits of agglomerative increases in socio-economic outputs (denser socioeconomic networks) are balanced with the costs of maintaining infrastructure networks and transporting goods, services, and individuals throughout a city. These considerations result, under population averaging, in urban scaling laws that describe how different urban quantities scale with city size, $N$, defined as the number of individuals living in a city~\cite{bettencourt2013origins,bettencourt2021introduction}. These empirically validated scaling laws have been found to hold for many cities across cultures and human history~\cite{bettencourt2013origins,bettencourt2021introduction,ortmanplos,ORTMAN201694,Ortmane1400066}.

For the average per-capita number of social interactions, $k$, the urban scaling law takes form of $k\sim N^\delta$, where $\delta=\frac{1}{6}$. Corrections to these exponent values due to growth rate fluctuations and other higher-order effects, provide the basis for a statistical theory of urban scaling \cite{bettencourt2021introduction, bettencourt2020urban}. Nevertheless, the simplest and most widely used form of this scaling law results from a mean-field approximation that individuals within a city interact with others homogeneously, without restrictions of group affiliation. Under such conditions, taking individuals to have an interaction cross section $a_0$ and a characteristic travel length $l$ per unit time, this approximation gives the average number of interactions for a large city ($N>>1$) as given by:
\begin{equation}
    k \sim \frac{a_0 l}{A}N
    \label{eq:scaling}
\end{equation}, where A is the area of the city's networks.

The scaling law for $k$ can be recovered from Equation \ref{eq:scaling} by substituting the scaling law for the area of the city's networks, $A \sim N^{1-\delta}$~\cite{bettencourt2013origins}, which is the result of self-consistently balancing the net benefits of socioeconomic interactions with costs of transportation (and housing) overbuilt urban spaces.

Importantly, $\frac{a_0l}{A}$ takes the role of a probability built out of the fraction of a city's area that individuals cover on average over a given time period. This is the average probability of interacting with all other individuals in the city, $N$. Thus, the total expected number of interactions for each individual (Equation \ref{eq:scaling}) is given by their probability of interacting, $\frac{a_0l}{A}$, multiplied by the number of individuals they could interact with, $N$.

Equation \ref{eq:scaling}, assumes that all individuals in the city are equally likely to interact ~\cite{bettencourt2013origins}. However, this assumption is unrealistic and can be relaxed by assuming that individuals belong to a number of distinct groups, which in turn have group-specific interaction rates. 

More specifically we model individuals in each of these groups as interacting preferentially with others of the same group, and with a lower probability with other groups. We define this relative reduction in out-group interactions by $1-h_g^{het}$, where $h_g^{het} \in [0,1]$ is the heterophobia of group $g$. Thus, $h_g^{het}=0.8$ means that individuals from group $g$ will only interact with 20\% of out-group members, on average. However, when individuals lose contact with other groups, they may compensate by having a higher rate of intra-group interactions, depending on the social context~\cite{mollenhorst2008social,skvoretz2013diversity}. We model this compensatory effect with a similar parameter, $h_g^{hom} \in [0,1]$, which is the homophily of group $g$ and sets the relative rate of intra-group interactions at $1+h_g^{hom}$. Though, $h_{g}^{het}$, and $h_{g}^{hom}$ are uncorrelated here by assumption, different social contexts may induce positive or negative correlations between $h_{g}^{het}$, and $h_{g}^{hom}$~\cite{mollenhorst2008social,skvoretz2013diversity}. In addition, individuals in a city have a limit on the number of interactions they can take part in during any given time period so that e.g., when all of an individual's interactions are within their own group they do not interact with other groups (complete outcome homophily and heterophobia). Taking this into account, future work might specify a budget for interactions that gives individuals the flexibility to trade between large numbers of less costly interactions (e.g., accessible within-group interactions) and fewer numbers of more costly but more rewarding interactions (e.g., to increase diversity and achieve superior group-level problem solving abilities~\cite{barkoczi2016social}).

It is important to note that here, $h_g^{hom}$ and $h_g^{het}$ are understood to represent outcome homophily and heterophobia and do not prescribe strong individual choices (i.e., strong individual preferences or avoidance of groups). In other words, $h_g^{hom}$ and $h_g^{het}$ are the result of observed network segregation of group $g$ that results from a combination of structural segregation and, possibly, individual preferences for certain groups. Though we expect relative increases in the rates of within-group interactions and relative decreases in the rate of between-group interactions, our model is agnostic to the direction of these effects (see Supplementary Text).

With these definitions, the average number of interactions for individuals in group $g$ is given by:
\begin{equation}
    k_g \sim \frac{a_0l}{A}[N_g\cdot(1+h_g^{hom}) + \sum_{j\neq g}N_j(1-h_g^{het})]
    \label{eq:group}
\end{equation}where $N_g$ is the population of focal group $g$. The total number of social interactions for all individuals in group $g$ is $k_g N_g$, on average. Therefore, the average number of social interactions for individuals in a observed segregated city $i$ with $G$ different groups, is $k_i\sim \frac{1}{N}\sum_{g=1}^G k_{g,i} N_{g,i}$. Here the average number of interactions for each group, $k_{g,i}$ and the size of each group, $N_{g,i}$ are specific to the observed city. This simplifies to (see Supplementary Text):
\begin{equation}
    k_i\sim k_0N_i^\delta\cdot(1-A_{i}^{het}+A_{i}^{hom})\cdot e^{\xi_i}
    \label{eq:hom}
\end{equation} with \begin{equation}
    A_i^{het}=\sum_{g=1}^G\sum_{j=g+1}^G \frac{N_{g,i}}{N_i} \frac{N_{j,i}}{N_i}(h_{g,i}^{het}+h_{j,i}^{het});\ \ \ \ A_i^{hom}=\sum_{g=1}^G(\frac{N_{g,i}}{N_i})^2 h_{g,i}^{hom}.
\end{equation} Here, $k_0$ is the scaling prefactor, $A_i^{het}$  is the heterophobia adjustment, $A_i^{hom}$ is the homophily adjustment, and $\xi_i$ are additional city specific effects~\cite{bettencourt2010urban}. The $A_i^{het}$, $A_i^{hom}$ are simply the averages of the coefficients $h_{g,i}^{het}, h_{g,i}^{hom}$, weighted by group sizes in each city. We see that  $1-A_i^{het}+A_i^{hom}$ gives a city's specific multiplicative adjustment to the scaling law. Note that $A_i^{het} \in [0,.25]$, while $A_i^{hom} \in [0,1]$ for a city with at least two groups so that, in this realization of the model, interactions are increased within-groups and decreased between-groups (see Supplementary Text). 

Thus, we expect that increased segregation between groups reduces social interactions (unless fully compensated for by increased within-group interactions), in line with previous research~\cite{ibarra1992homophily}. In addition, $A_i^{het}$ depends on the relative sizes of groups and has the largest impact when all groups are of equal size (see Supplementary Text), matching previous investigations of homophily and heterophobia in small networks~\cite{oliveira2022group}.  

The final step is to consider how interactions between city inhabitants translate to economic outcomes. In the derivation of urban scaling laws for economic outputs, interactions over various types (friendship, employment, acquaintance, etc) can couple to economic outputs either positively or negatively, and with varying strengths over the different types~\cite{bettencourt2013origins}. Similarly, interactions between- or within-groups can couple deferentially to social and economic outputs, so that, for instance, within-group interactions might be more productive for social outputs, but less productive for creative outputs~\cite{barkoczi2016social}. For simplicity, here, we assume that between and within group interactions do not couple to economic outputs differently, so that economic outputs are directly proportional to the social interactions specified by Equation \ref{eq:hom}~\cite{bettencourt2013origins,bettencourt2021introduction}.

\subsection*{Empirical tests of the homophily and heterophobia adjustments}

We next sought to test the empirical validity of the homophily and heterophobia adjustments regarding self-reported race in U.S. cities. $A_i^{het}$ and $A_i^{hom}$ were calculated from racial demographic estimates in cities collected by the U.S. census for each year between 2010-2020 (see Materials and Methods). \begin{figure}[bpht!]
     \centering
     \includegraphics{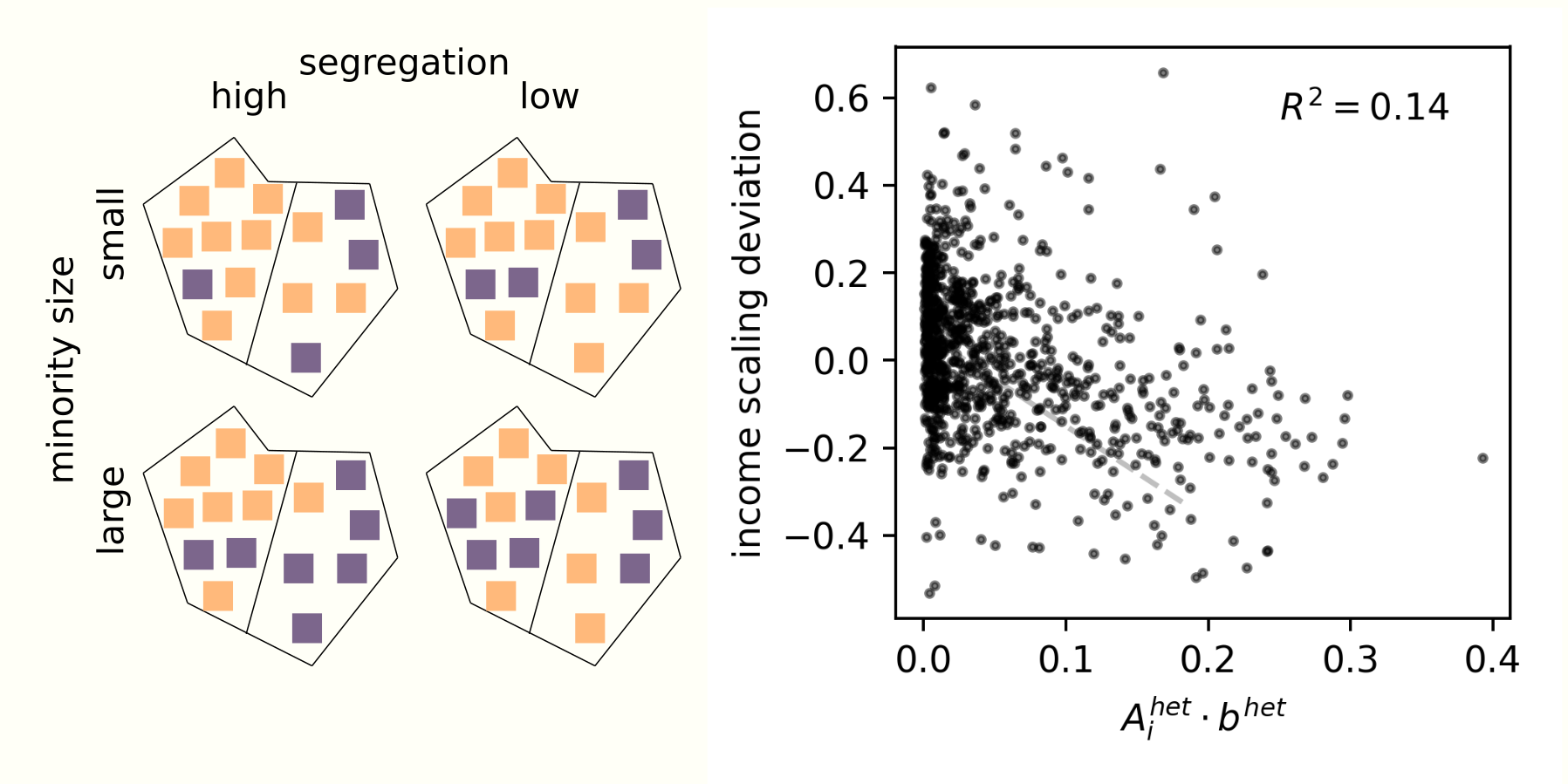}
     \caption{Left: Measures of citywide segregation depend on the relative size of the groups and their spatial concentration. Starting at the top left and going clockwise, $A_i^{het} \sim \sum\frac{N_{g,i}}{N_i}\frac{N_{j,i}}{N_i}(s_{g,i}+s_{j,i})$ takes on values of 0.92, 0.98, 0.98, and, 0.85. Here, $s_{g,i}$ is the level of residential segregation of group $g$ in city $i$.  The large minority, high segregation city in the lower left has the smallest $A_i^{het}$. Top: a city where the majority makes up 73\% of the population and the average segregation is 20\% (left) or 5.5\% (right). Bottom: a city where the majority makes up 53\% of the population and the average segregation is 30\% (left) or 3\% (right).   Right: The relationship between urban scaling law deviations for median income and $A_i^{het}\cdot b^{het}$ in U.S. cities in 2020.}
     \label{fig:one}
 \end{figure}
 Racial segregation for each city and group was calculated as the average difference between neighborhood proportions of residents in each racial group and the city-wide mean. Explicitly, $s_{g,i} = \frac{1}{M}\sum_{m=1}^{M}|N_{g,m,i}/N_{m,i} - N_{g,i}/N_i|$, where $m$ indexes all of the neighborhoods in a city and $N_{g,m,i}/N_{m,i}$ is the proportion of residents in racial group $g$ for neighborhood $m$ in city $i$ (see Figure \ref{fig:one}; we note that residential segregation may only capture some of the overall social contact segregation that occurs in cities~\cite{tammaru2016relations,priest2014patterns,tucker2021tweets}). For example, in a city with two groups where the majority is 80\% of the population, one neighborhood might have 90\% of its population from the majority group while another neighborhood is 20\% majority group; in this case, the majority group would have a segregation value of $0.5\cdot(|0.9-0.8|+|0.2-0.8|) = 0.35$. Since we expect increased segregation, on average, to lead to increased homophily and heterophobia values, though possibly with different strengths, we modeled these values as linearly related to the empirical segregation values: $h_i^\cdot \sim b^\cdot*s_i$, where $b^\cdot$ determines the strength of coupling between residential racial segregation and heterophobia. In order to ensure that our results were not sensitive to the choice of segregation measure we repeated analyses with four additional segregation measures (see Materials and Methods).

 We performed these analyses for two measures of economic productivity, median income and gross domestic product (GDP), in order to assess the impacts of segregation on individual and overall economic productivity in U.S. cities, respectively. Though median income and GDP are correlated (Spearman r $\sim 0.55$, see Supplementary Table \ref{tab:gdpIncCors}), the ability of individuals to garner higher wages and of businesses to generate high economic outputs are not commensurate~\cite{nolan2019gdp}.
 
 We chose to conduct our analyses at the level of census tracts for the U.S., which are small spatial units with approximately 4,000 inhabitants (results were similar when smaller spatial units of census block groups were used, see Supplementary Text). Analyses were conducted within functional cities (integrated commuting areas), defined as combined statistical areas by the U.S. Office of Budget and Management~\cite{office2021recommendations}. These are functional definitions of cities that capture where people live, socialize, and work~\cite{bettencourt2021introduction,stier2022reply}. 


 These analyses revealed the variation in scaling deviations explained by homophily and heterophobia and the empirical strength, $b^{het,hom}$, of $A_i^{het}$ and $A_i^{hom}$, i.e., the degree to which segregation impacts interactions between and within groups, respectively. The results demonstrate that scaling deviations for income are significantly predicted by $A_i^{het}$ across all years (Figure \ref{fig:one}, Supplementary Table \ref{tab:incHet}), but not by $A_i^{hom}$ (Supplementary Table \ref{tab:incHet}). Though $A_i^{het}$ and $A_i^{hom}$ are strongly correlated ($\bar{r}\sim 0.77$, range $[0.75, 0.78]$ across years), the variance inflation factor is relatively low when using centered versions of these variables ($VIF < 2.58$ for all years) and the Akaike information criterion (AIC) of models with only $A_i^{het}$ is always lower than models with only $A_i^{hom}$ (average $\Delta$AIC $=-44$, range $[-54,-37]$).
 
 In addition, $A_i^{het}$ explains a maximum of 13.7\% of the variance in scaling deviations for income in 2020 (Figure \ref{fig:one}, Supplementary Table \ref{tab:inc}). In contrast, $A_i^{het}$ is only significantly predictive of GDP scaling deviations after 2014 where it explains a maximum 1.3\% of the variance in 2020 (Supplementary Table \ref{tab:gdp}. Results were similar when alternate measures of residential racial segregation were used (see Materials and Methods, Supplementary Tables \ref{tab:segINC}-\ref{tab:expGDP}). These results suggest that heterophobia has a stronger effect on economic productivity than homophily and that it is more important for individual outcomes than for the whole of cities' economies.
 
 One reason for these differences might be that the types of interactions that racial segregation curtails are more important for labor opportunities and associated opportunities for securing higher wages and incomes than they are for the overall productivity of firms in an urban economy~\cite{alabdulkareem2018unpacking}. These differences could be operationalized in future work through differential coupling of various modes of between- and within-group interactions to various types of economic outputs at different scales of organization. 
 
  \begin{figure}[bpht!]
     \centering
     \includegraphics{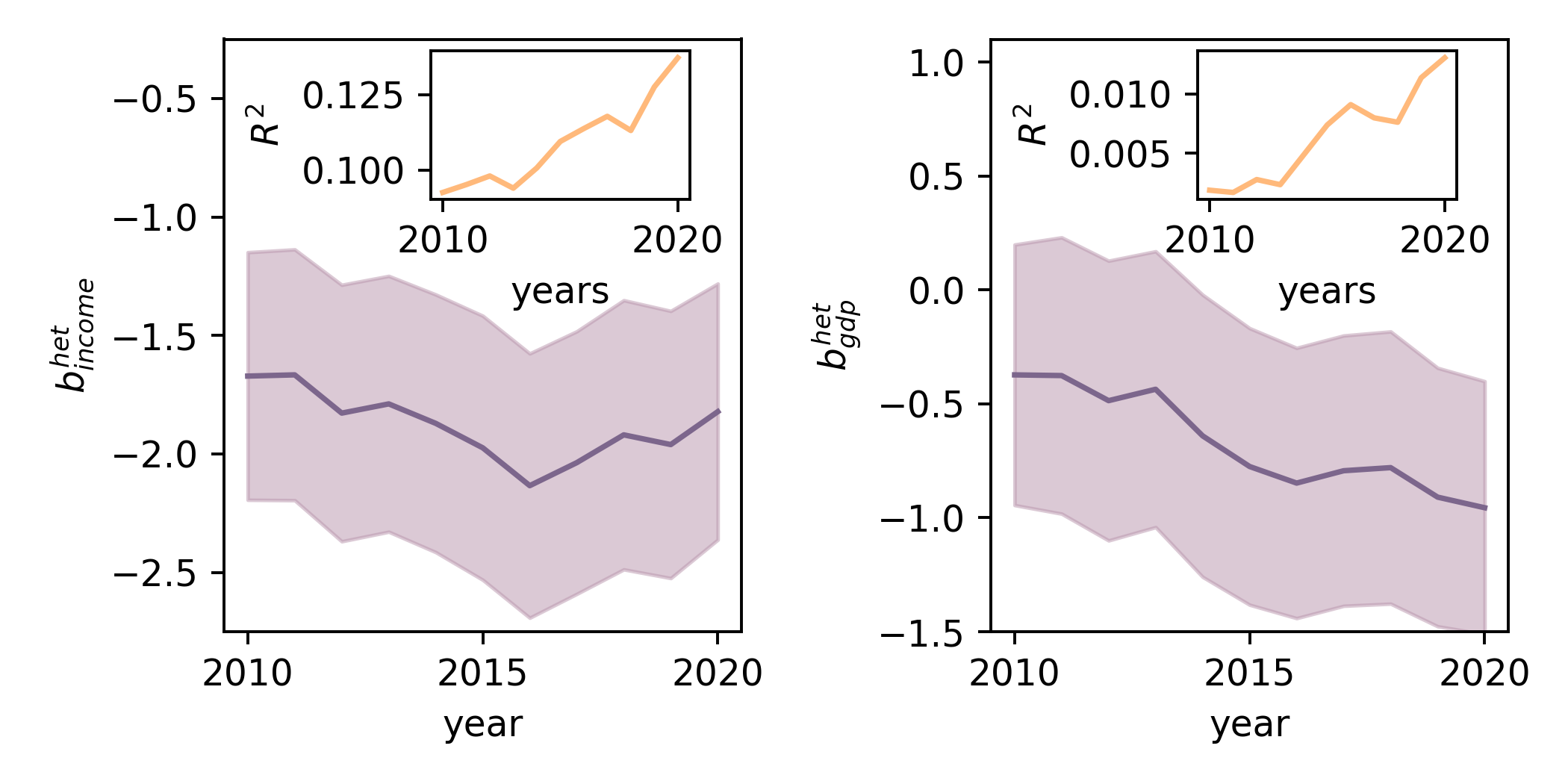}
     \caption{Changes in the empirical strength of the heterophobia adjustment over time. Insets show $R^2$ values for the OLS regression models. Shaded regions show the 95\% confidence interval for $b^{het}$, i.e. the strength of the relationship between $h_g^{het}$ and $s_g$. The different values of $b^{het}$ are parameters in the OLS regression models. Left: median income. Right: gross domestic product (GDP).}
     \label{fig:two}
 \end{figure}
 
 Finally, we observe that the degree to which residential segregation is associated with economic productivity increased between 2010 and 2020 (Figure \ref{fig:two}). In particular, for both median income and GDP, the explained variance of scaling deviations, $R^2$, increased systematically between 2010 and 2020. This happened while the average value of $A_i^{het}$ stayed relatively constant (Supplementary Figure \ref{fig:Sch}). Thus, though observed racial segregation in U.S. cities was relatively constant during the 2010s, the degree to which segregation accounted for lower incomes and GDP, relative to city size, likely increased.
 
\section*{Discussion}
Our development of urban scaling relations to account for racial segregation effects suggests that heterogeneous network structure in cities can meaningfully impact their economic outputs. The adjustment to urban scaling laws we derived suggests that segregation reduces economic outputs with strength depending on the size of different groups and their levels of heterophobia~\cite{lee2019homophily}. Our empirical findings support the hypothesis that this is indeed the case. For example, in 2020, the New Orleans, Louisiana Metropolitan Area had a median income of approximately \$54,400. The urban scaling law predicts a median income of approximately \$65,000 from city size alone, while accounting for segregation brings the prediction to approximately \$56,500. 

In general, interactions beyond the residence, in shared public spaces and in workplace environments, are also likely relevant to economic outputs.  Consequently, it is important that future work model heterogeneous, segregated ambient mixing in these environments  and evaluate it alongside residential segregation. Such considerations are particularly important for understanding, more generally, how differences in structured group interactions lead to more or less productive cities.

Moreover, our observation that the influence of heterophobic preferences on economic outputs is increasing over time suggests that additional factors are accelerating the detrimental effects of segregation in United States cities. These might include factors like peer influence and behavioral norms~\cite{jackson2021inequality} which can interact with heterophobia to exacerbate induced inequalities. 

Why do individual economic outcomes (e.g., income) show stronger associations with racial segregation than overall economic output? How does racial segregation effect other well-described urban social behaviors such as mental health~\cite{stier2021evidence} and crime~\cite{bettencourt2010urban}? How do other forms of segregation of interactions, via e.g., politics and wealth, impact the various economic and social outputs? Such questions are important for future research and require further development of theoretical models and empirical investigations along the lines developed here. It is important that this research is performed, at least in part, in the context of cities, which while housing enormous human diversity often fail to make the most of their latent socioeconomic potential. Urban environments are now home to the majority of the world's population and account for a disproportionate fraction of global economic and intellectual productivity; as a result, a better understanding of the sources of inequality, and successful integration of diversity in cities is crucial to building more equitable and inclusive societies.

\section*{Materials and Methods}
\subsection*{U.S. Census and Economic Data}
All census data is publicly available and was downloaded from \url{data.census.gov}. Five year racial demographic estimates for U.S. census tracts and census block groups were downloaded from table \textit{B02001}. Homophily values were calculated across the six racial groups provided in these tables: White, Black, Native American/Native Alaskan, Asian, Hawaiian/Pacific Islander, and Other. Five-year population estimates for U.S. cities defined as combined statistical areas (CBSAs) were downloaded from table \textit{B01003}. We note that these demographic estimates are only available for block groups from 2013 onward, but are available for census tracts from 2010 onward. Five-year median income estimates for U.S. cities were downloaded from table \textit{B19013}. Gross domestic product data for CBSAs is publicly available from the Bureau of Economic Analysis and was downloaded from \url{https://apps.bea.gov/iTable/index_regional.cfm}. In order to map between census tracts and CBSAs, delineation files for 2020 were downloaded from the United States Office of Budget and Management from \url{https://www.census.gov/programs-surveys/metro-micro/about/delineation-files.html}.

\subsection*{Association Between Scaling Deviations and $A_i^{het}$ and $A_i^{hom}$}
In order to determine the association between $A_i^{het}$ and $A_i^{hom}$ and measures of economic productivity, the scaling relationship between economic measures and city population has to be removed. However, we first recognize that our measured segregation values, though $\in[0,1]$ are not true probabilities. However, we can expect $h_{g,i}^{hom}$ and $h_{g,i}^{het}$ to be proportional to the empirical segregation values. Explicitly, \begin{equation}
    h_{g,i}^\cdot = h^\cdot + b^\cdot s_{g,i}
\end{equation}, where $h^\cdot$ is the base probability, $b^\cdot$ a scaling factor, and $s_{g,i}$ the empirical segregation values for group $g$ in city $i$. We have, for simplicity, assumed a linear relationship so that the hetrophobia and homophily values increase linearly with the degree of segregation.

$A_i^{het}$ and $A_i^{hom}$ then become:
\begin{equation}
    A_i^{het} = \sum_{g=1}^G\sum_{j=g+1}^G\frac{N_{g,i}}{N_i}\frac{N_{j,i}}{N_i}[2\cdot h^{het}+  b^{het}\cdot(s_{g,i}+s_{g,j})];\ \ \ \ A_i^{hom} =\sum_{g=1}^G(\frac{N_{g,i}}{N_i})^2\cdot[h^{hom} + b^{hom}\cdot s_{g,i})]
\end{equation} when the combined homopohily and heterophobia effects are small we can approximate this by ($ln(1+x) \sim x$ when $x<<1$): \begin{equation}
    ln(k_i)\simeq \delta \cdot ln(N_i)-\sum_{g=1}^G\sum_{j=g+1}^G\frac{N_{g,i}}{N_i}\frac{N_{j,i}}{N_i}[2\cdot h^{het} + b^{het}\cdot(s_{g,i}+s_{g,j})]+\sum_{g=1}^G(\frac{N_{g,i}}{N_i})^2\cdot[h^{hom} + b^{hom}\cdot s_{g,i}]+\xi_i
    \label{eq:reg_start}
\end{equation} since our primary interest is on the effects of segregation, we can write this as two regression equations: \begin{equation}
    ln(k_i)\sim ln(C) + \beta \cdot ln(N_i) + \epsilon_i
    \label{eq:FirstReg}
\end{equation} and,
\begin{equation}
    \epsilon_i \sim D -b^{het}\cdot \sum_{g=1}^G\sum_{j=g+1}^G\frac{N_{g,i}}{N_i}\frac{N_{j,i}}{N_i}(s_{g,i}+s_{j,i})+b^{hom}\cdot \sum_{g=1}^G(\frac{N_{g,i}}{N_i})^2\cdot s_{g,i}+ \xi_i
    \label{eq:SecondReg}
\end{equation} where we have included the additional city specific terms of Equation \ref{eq:reg_start} in the residuals $
\xi_i$ which are the same city specific effects from Equation \ref{eq:hom}. Here, $C$ and $D$ are city size independent constants, i.e., scaling prefactors. In addition, $\beta$ is the scaling exponent estimate which we expect to take on the value of $\delta=\frac{1}{6}$.

We estimated the regression for Equation \ref{eq:FirstReg} by OLS first and then used the scaling deviations, $\epsilon_i$, from those regressions to estimate the influence of homophily and heterophobia by OLS via Equation \ref{eq:SecondReg}. In order to exclude outlier cities that significantly deviate from the scaling law, cities with $|\epsilon|>3.09\sqrt{Var(\epsilon)}$, i.e., beyond the 99.9\%th percentile of the normal distribution of the standard deviation of $\epsilon$ were excluded for each year. Results are similar when outliers are not excluded (Supplementary Tables \ref{tab:outlierGDP} \& \ref{tab:outlierInc}).

\subsection*{Alternate Measures of Residential Segregation}
 In order to ensure that the results were not sensitive to a specific segregation measure we repeated analyses with three additional measures of residential segregation~\cite{white1986segregation}. These included the normalized segregation index:
 \begin{equation}
     D_{g,i} = \frac{\sum_m |\frac{N_{g,m,i}}{N_{m,i}} - \frac{N_{g,i}}{N_i}| \cdot N_{m,i}}{2\cdot N_i \cdot \frac{N_{g,i}}{N_i} \cdot (1-\frac{N_{g,i}}{N_i})}
 \end{equation} the Gini Coefficient:
\begin{equation}
     gini_{g,i} = \frac{\sum_m \sum_l |\frac{N_{g,m,i}}{N_{m,i}} - \frac{N_{g,l,i}}{N_{l,i}}| \cdot N_{m,i} \cdot N_{l,i}}{2\cdot N_i^2 \cdot \frac{N_{g,i}}{N_i} \cdot (1-\frac{N_{g,i}}{N_i})}
 \end{equation} and the exposure $B_{gg}$ index, also known as the correlation ratio (CR or $\eta^2$) or the mean squared deviation: \begin{equation}
     \eta^2_{g,i} = \frac{\sum_m N_{g,m,i}^2}{N_{g,i}\cdot (1-\frac{N_{g,i}}{N_i})} - \frac{\frac{N_{g,i}}{N_i}}{1-\frac{N_{g,i}}{N_i}}
 \end{equation}

\noindent The construction of measures of segregation based more directly on interactions, beyond the composition of residential neighborhoods, is also likely important and will be pursued in future work.
 
\bibliography{scibib}

\bibliographystyle{Science}

\section*{Acknowledgments}
The authors declare that they have no competing interests. All data needed to evaluate the conclusions in the paper are present in the paper or are publicly available, the Supplementary Materials, or are publicly available.

Author contributions: A.J.S. and F.K designed research; A.J.S. performed research; M.G.B supervised research; and A.J.S., S.S, F.K., L.M.A.B, and M.G.B. wrote the paper.
\\

\noindent This work is partially supported by NSF-2106013, and S\&CC-1952050.
\newpage


\newpage
\clearpage
\section*{Supplementary Text}
\setcounter{equation}{0}

\subsection*{Derivation of the homophily and hetrophobia adjustments}
Here we expand on the derivation of the homophily and heterophobia adjustments in the main text. We start with Equation \ref{eq:group} of the main text which gives the average number of social interactions per individual in group $g$. From this we can write down the average number of social interactions per individual for the entire city as a sum over the different groups, $G$. For simplicity we have dropped the residual term $e^{\xi_i}$:
\begin{equation}
    k_i \sim \frac{a_0l}{A_iN_i}[\sum_{g=1}^G(N_{g,i}(1+h^{hom}_{g,i})+\sum_{j\neq g}N_{j,i}(1-h^{het}_{g,i}))\cdot N_{g,i}]
\end{equation} multiplying through by $N_{g,i}$ we have:
\begin{equation}
    k_i \sim \frac{a_0l}{A_iN_i}[\sum_{g=1}^G(N_{g,i}^2)+\sum_{g=1}^G(N_{g,i}^2h^{hom}_{g,i})+\sum_{g=1:G}\sum_{j\neq g}N_{g,i}N_{g,j} - \sum_{g=1:G}\sum_{j\neq g}N_{g,i}N_{j,i}h_{g,i}^{het}]
\end{equation} since the third term in the brackets gives two copies of each $N_{g,i}N_{j,i}$ term, we can then write:
\begin{equation}
    k_i \sim \frac{a_0l}{A_iN_i}[(\sum_{g=1}^GN_{g,i})^2 - \sum_{g=1}^G\sum_{j=g+1}^GN_{g,i}N_{j,i}(h_{g,i}^{het}+h_{j,i}^{het})+\sum_{g=1}^G(N_{g,i}^2h^{hom}_{g,i})]
\end{equation} and finally, we divide and multiply the second term by $N_i^2$ and arrive back at Equation \ref{eq:hom} of the main text:
\begin{equation}
    k_i \sim \frac{a_0l}{A_iN_i}[N_i^2 - N_i^2\sum_{g=1}^G\sum_{j=g+1}^G\frac{N_{g,i}}{N_i}\frac{N_{j,i}}{N_i}(h_{g,i}^{het}+h_{j,i}^{het})+N_i^2\sum_{g=1}^G(\frac{N_{g,i}}{N_i})^2h^{hom}_{g,i}]
    \label{eq:suphom}
\end{equation} equivalently:
\begin{equation}
    k_i \sim \frac{a_0lN_i}{A_i}[1 - A_i^{het}+A_i^{hom}]
\end{equation}
Note that there are no homophily and heterophobia effects Equation \ref{eq:suphom} becomes:\begin{equation}
    k_i \sim \frac{a_0l}{A_iN_i}[N_i^2] = \frac{a_0lN_i}{A_i}
\end{equation} and we recover the typical scaling law.

In the main text, we made a comment that $A_i^{het}$ is always less than $0.5$ and $\geq 0$. This can be easily seen from the fact that $A_i^{het}$ is at its smallest when there are only two groups and those groups are equally balanced. In this case, $A_i^{het}$ takes on a value of $0.5\cdot 0.5 \cdot(1+1)$ when the two groups are completely heterophobic and is smaller for groups with less heterophobia, imbalanced proportions of the population, or for more than two groups. In general, for $G>2$ groups, $A_i^{het}$ is at most $2/G^2\cdot \genfrac(){0pt}{}{G}{2}=(G-1)/G$ which occurs when the heterophobia values for all groups are $1$ and groups are equal in size. Similarly, $A_i^{hom}$ is at minimum $1/G$ and this occurs when homophily values for all groups are $1$ and groups are equal in size.

As a note, in general, heterophobia values need not be the same across all groups. In particular, we can define a city specific matrix with entries $\in[-1,1]$ which specifies the degree to which individuals avoid or preferentially interact with individuals from the same or other groups:\begin{equation}
    H_i = \left[ {\begin{array}{cccc}
    h_{11i} & h_{12i} & \cdots & h_{1Gi}\\
    h_{21i} & h_{22i} & \cdots & h_{2Gi}\\
    \vdots & \vdots & \ddots & \vdots\\
    h_{G1i} & h_{G2i} & \cdots & h_{GGi}\\
  \end{array} } \right]
\end{equation} The relative rate of interactions between any two groups (or within a group) is given by $1+H_i$.  In the main text we assumed homophily and heterophobia, i.e., that the diagonal elements of $H$ are positive and the off diagonal elements of $H$ are negative. However, in general, homophobia and heterophilly may also be present so that the entries of $H$ can take on positive or negative values. In this case, positive values correspond to homophily or heterophily and negative values correspond to homophobia or heterophobia. 

Our specific constraints on $H$ in the models presented in the main text were that: (1) homophily is specified by the diagonal entries of $H_i$, $h^{hom}_{g,i}=h_{ggi}$, which we assume to be positive, and (2) that each group avoids all other groups equally, so that there are repeated off diagonal entries and that all of the off diagonal entries are negative. Specifically, that $h^{het}_{g,i}=h_{gji}$ for all $j\neq g$.

This notation allows for an alternative notation for Equation \ref{eq:suphom}:
\begin{equation}
    k_i \sim \frac{a_0l}{A_iN_i}[tr(\mathbf{N_i}H_i\mathbf{N_i}) + tr([1-\mathbf{I}]\mathbf{N_i}H_i\mathbf{N_i}]
\end{equation} where $\mathbf{N_i}$ is the diagonal matrix of group sizes and $I$ is the identity matrix. In other words, between-group interactions are given by the sum of off diagonal elements of $\mathbf{N_i}H\mathbf{N_i}$ and within-group interactions are given by the sum of diagonal elements.

\subsection*{Census Block Group Analysis}
All of the analyses described in the main text were also conducted with census block groups instead of census tracts. In contrast to census tracts which contain, on average, 4,000 individuals, census block groups contain 1,500 individuals on average. A similar pattern of results was found for census block groups.

\begin{table}[bpht!]
\centering
\caption{Fits of calculated heterophobia adjustments to median income scaling deviations by year. $b^{het}$ determines the strength of the coupling between economic productivity and residential segregation by controlling levels of heterophobia associated with residential segregation.}
\begin{tabular}{lll}

year & $b^{het}$ 95\% CI & $R^{2}$ \\
\hline\hline
2013 & [1.03, 1.60] &    0.09 \\
2014 & [1.10, 1.67] &    0.10 \\
2015 & [1.25, 1.84] &    0.11 \\
2016 & [1.35, 1.93] &    0.13 \\
2017 & [1.39, 1.96] &    0.13 \\
2018 & [1.37, 1.96] &    0.13 \\
2019 & [1.46, 2.04] &    0.14 \\
2020 & [1.46, 2.04] &    0.14 \\

\end{tabular}
\end{table}
\begin{table}[bpht!]
\centering
\caption{Fits of calculated heterophobia adjustments to GDP scaling deviations by year.}
\begin{tabular}{lll}

year & $b^{het}$ 95\% CI & $R^{2}$ \\
\hline\hline
2013 & [-0.12, 0.93] & 0.00283 \\
2014 &  [0.01, 1.06] & 0.00476 \\
2015 &  [0.07, 1.11] & 0.00597 \\
2016 &  [0.17, 1.18] & 0.00807 \\
2017 &  [0.19, 1.20] & 0.00873 \\
2018 &  [0.18, 1.19] & 0.00831 \\
2019 &  [0.27, 1.23] & 0.01072 \\
2020 &  [0.26, 1.21] & 0.01076 \\

\end{tabular}
\end{table}

\newpage
\clearpage
\section*{Supplementary Figures}
\setcounter{figure}{0}
\renewcommand{\figurename}{Supplementary Figure}

 \begin{figure}[bpht!]
     \centering
     \includegraphics{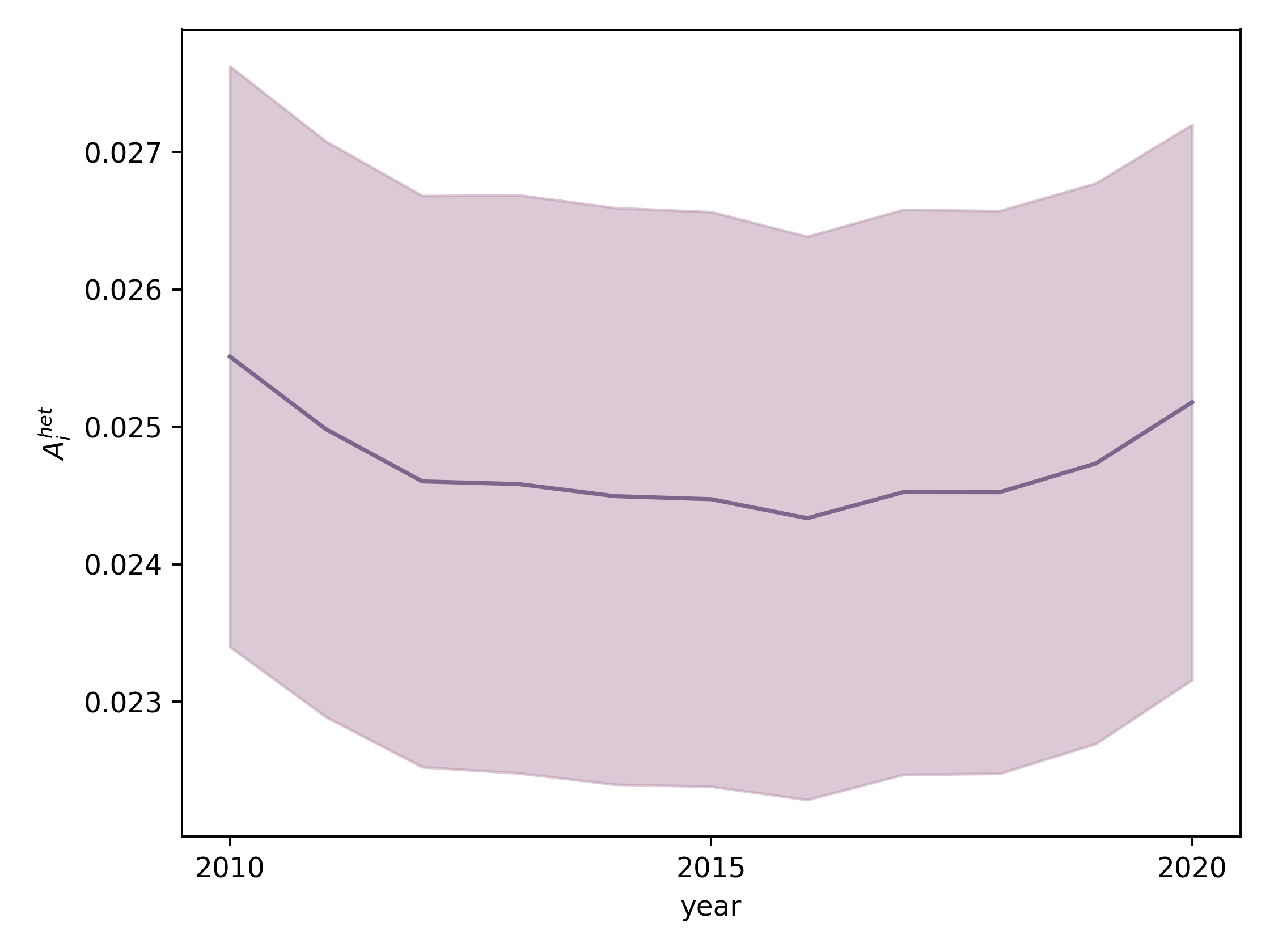}
     \caption{Mean $A_i^{het}$ across cities over time. The shaded region represents the 95\% interval of the standard error of the mean.}
     \label{fig:Sch}
 \end{figure}

\newpage
\clearpage
\section*{Supplementary Tables}
\renewcommand{\tablename}{Supplementary Table}

\begin{table}[bpht!]
\centering
\caption{Spearman Rank Order Correlation between Median Income and GDP by year.}
\label{tab:gdpIncCors}
\begin{tabular}{lll}

year & $r\_s$ &  p-value \\
\hline\hline
2010 &  0.56 & 1.77e-74 \\
2011 &  0.56 & 3.38e-72 \\
2012 &  0.55 & 2.52e-71 \\
2013 &  0.54 & 2.08e-66 \\
2014 &  0.53 & 6.96e-64 \\
2015 &  0.52 & 1.19e-60 \\
2016 &  0.52 & 1.34e-62 \\
2017 &  0.53 & 7.85e-65 \\
2018 &  0.54 & 1.97e-67 \\
2019 &  0.55 & 8.83e-71 \\
2020 &  0.56 & 5.50e-73 \\

\end{tabular}
\end{table}

\begin{table}[bpht!]
\centering
\caption{Fits of calculated heterophobia corrections to median income scaling deviations by year.}
\label{tab:incHet}
\begin{tabular}{llll}

year &   $b^{het}$ 95\% CI & $R^{2}$ &   n \\
\hline\hline
2010 & [-1.88, -1.23] &    0.09 & 872 \\
2011 & [-1.94, -1.28] &    0.10 & 872 \\
2012 & [-2.02, -1.34] &    0.10 & 873 \\
2013 & [-1.96, -1.28] &    0.09 & 851 \\
2014 & [-2.06, -1.37] &    0.10 & 852 \\
2015 & [-2.20, -1.49] &    0.11 & 852 \\
2016 & [-2.28, -1.56] &    0.11 & 865 \\
2017 & [-2.31, -1.60] &    0.12 & 865 \\
2018 & [-2.31, -1.58] &    0.11 & 866 \\
2019 & [-2.44, -1.71] &    0.13 & 874 \\
2020 & [-2.49, -1.77] &    0.13 & 874 \\

\end{tabular}
\end{table}

\begin{table}[bpht!]
\centering
\caption{Fits of calculated homophily and heterophobia corrections to median income scaling deviations by year.}
\label{tab:inc}
\begin{tabular}{lllll}

year &   $b^{het}$ 95\% CI &  $b^{hom}$ 95\% CI & $R^{2}$ &   n \\
\hline\hline
2010 & [-2.19, -1.15] & [-0.35, 0.63] &    0.09 & 872 \\
2011 & [-2.20, -1.14] & [-0.43, 0.57] &    0.10 & 872 \\
2012 & [-2.37, -1.29] & [-0.33, 0.69] &    0.10 & 873 \\
2013 & [-2.33, -1.25] & [-0.31, 0.71] &    0.09 & 851 \\
2014 & [-2.41, -1.33] & [-0.32, 0.71] &    0.10 & 852 \\
2015 & [-2.53, -1.42] & [-0.37, 0.69] &    0.11 & 852 \\
2016 & [-2.69, -1.58] & [-0.26, 0.79] &    0.11 & 865 \\
2017 & [-2.59, -1.48] & [-0.43, 0.63] &    0.12 & 865 \\
2018 & [-2.49, -1.35] & [-0.58, 0.51] &    0.11 & 866 \\
2019 & [-2.52, -1.40] & [-0.70, 0.41] &    0.13 & 874 \\
2020 & [-2.36, -1.28] & [-1.00, 0.14] &    0.14 & 874 \\

\end{tabular}
\end{table}

\begin{table}
\centering
\caption{Fits of calculated heterophobia corrections to GDP scaling deviations by year.}
\label{tab:gdp}
\begin{tabular}{llll}

year &   $b^{het}$ 95\% CI & $R^{2}$ &   n \\
\hline \hline
2010 &  [-0.94, 0.20] & 0.00191 & 861 \\
2011 &  [-0.98, 0.23] & 0.00172 & 863 \\
2012 &  [-1.10, 0.13] & 0.00280 & 862 \\
2013 &  [-1.04, 0.17] & 0.00237 & 843 \\
2014 & [-1.26, -0.02] & 0.00488 & 844 \\
2015 & [-1.38, -0.17] & 0.00738 & 847 \\
2016 & [-1.44, -0.25] & 0.00910 & 860 \\
2017 & [-1.39, -0.20] & 0.00798 & 859 \\
2018 & [-1.38, -0.18] & 0.00762 & 858 \\
2019 & [-1.48, -0.34] & 0.01135 & 865 \\
2020 & [-1.51, -0.40] & 0.01303 & 868 \\

\end{tabular}
\end{table}

\begin{table}
\centering
\caption{Fits of calculated homophily and heterophobia corrections to median income scaling deviations by year using the segregation index.}
\label{tab:segINC}
\begin{tabular}{lllll}

year &   $b^{het}$ 95\% CI &  $b^{hom}$ 95\% CI & $R^{2}$ &   n \\
\hline \hline
2010 & [-0.76, -0.45] & [-0.06, 0.17] &    0.07 & 872 \\
2011 & [-0.79, -0.47] & [-0.07, 0.17] &    0.07 & 872 \\
2012 & [-0.82, -0.50] & [-0.06, 0.19] &    0.07 & 873 \\
2013 & [-0.79, -0.46] & [-0.07, 0.18] &    0.06 & 851 \\
2014 & [-0.84, -0.51] & [-0.08, 0.17] &    0.07 & 852 \\
2015 & [-0.90, -0.56] & [-0.10, 0.15] &    0.08 & 852 \\
2016 & [-0.94, -0.59] & [-0.04, 0.23] &    0.08 & 865 \\
2017 & [-0.96, -0.62] & [-0.09, 0.18] &    0.08 & 865 \\
2018 & [-0.96, -0.61] & [-0.14, 0.14] &    0.08 & 866 \\
2019 & [-1.02, -0.67] & [-0.17, 0.11] &    0.09 & 874 \\
2020 & [-1.05, -0.70] & [-0.24, 0.04] &    0.10 & 874 \\

\end{tabular}
\end{table}

\begin{table}
\centering
\caption{Fits of calculated heterophobia corrections to GDP scaling deviations by year using the segregation index.}
\label{tab:segGDP}
\begin{tabular}{llll}

year &   $\pi$ 95\% CI & $R^{2}$ &   n \\
\hline \hline
2010 &  [-0.30, 0.24] & 0.00006 & 861 \\
2011 &  [-0.32, 0.25] & 0.00007 & 863 \\
2012 &  [-0.37, 0.20] & 0.00039 & 862 \\
2013 &  [-0.34, 0.23] & 0.00015 & 843 \\
2014 &  [-0.46, 0.13] & 0.00144 & 844 \\
2015 &  [-0.55, 0.03] & 0.00366 & 847 \\
2016 & [-0.58, -0.02] & 0.00508 & 860 \\
2017 & [-0.57, -0.00] & 0.00458 & 859 \\
2018 &  [-0.55, 0.02] & 0.00375 & 858 \\
2019 & [-0.57, -0.03] & 0.00533 & 865 \\
2020 & [-0.60, -0.07] & 0.00706 & 868 \\

\end{tabular}
\end{table}

\begin{table}
\centering
\caption{Fits of calculated homophily and heterophobia corrections to median income scaling deviations by year using the gini coefficient.}
\label{tab:giniINC}
\begin{tabular}{lllll}

year &   $b^{het}$ 95\% CI &  $b^{hom}$ 95\% CI & $R^{2}$ &   n \\
\hline \hline
2010 & [-0.59, -0.35] &  [-0.06, 0.12] &    0.06 & 872 \\
2011 & [-0.61, -0.37] &  [-0.08, 0.11] &    0.07 & 872 \\
2012 & [-0.63, -0.38] &  [-0.06, 0.13] &    0.07 & 873 \\
2013 & [-0.61, -0.36] &  [-0.07, 0.12] &    0.06 & 851 \\
2014 & [-0.64, -0.39] &  [-0.08, 0.13] &    0.07 & 852 \\
2015 & [-0.69, -0.43] &  [-0.10, 0.11] &    0.07 & 852 \\
2016 & [-0.72, -0.45] &  [-0.05, 0.16] &    0.08 & 865 \\
2017 & [-0.74, -0.47] &  [-0.08, 0.13] &    0.08 & 865 \\
2018 & [-0.74, -0.47] &  [-0.14, 0.08] &    0.08 & 866 \\
2019 & [-0.79, -0.52] &  [-0.16, 0.06] &    0.09 & 874 \\
2020 & [-0.81, -0.55] & [-0.23, -0.00] &    0.11 & 874 \\
\end{tabular}
\end{table}

\begin{table}
\centering
\caption{Fits of calculated heterophobia corrections to GDP scaling deviations by year using the gini coefficient.}
\label{tab:giniGDP}
\begin{tabular}{llll}

year &   $\pi$ 95\% CI & $R^{2}$ &   n \\
\hline \hline
2010 &  [-0.22, 0.20] & 0.00001 & 861 \\
2011 &  [-0.24, 0.20] & 0.00004 & 863 \\
2012 &  [-0.28, 0.17] & 0.00029 & 862 \\
2013 &  [-0.25, 0.19] & 0.00007 & 843 \\
2014 &  [-0.34, 0.12] & 0.00111 & 844 \\
2015 &  [-0.40, 0.05] & 0.00279 & 847 \\
2016 &  [-0.43, 0.01] & 0.00425 & 860 \\
2017 &  [-0.42, 0.02] & 0.00380 & 859 \\
2018 &  [-0.41, 0.03] & 0.00335 & 858 \\
2019 & [-0.43, -0.01] & 0.00473 & 865 \\
2020 & [-0.45, -0.05] & 0.00692 & 868 \\
\end{tabular}
\end{table}

\begin{table}
\centering
\caption{Fits of calculated homophily and heterophobia corrections to median income scaling deviations by year using the exposure index.}
\label{tab:expINC}
\begin{tabular}{lllll}

year &   $b^{het}$ 95\% CI &  $b^{hom}$ 95\% CI & $R^{2}$ &   n \\
\hline \hline
2010 & [-1.43, -0.70] & [-0.09, 0.44] &    0.06 & 872 \\
2011 & [-1.47, -0.72] & [-0.12, 0.42] &    0.07 & 872 \\
2012 & [-1.57, -0.80] & [-0.08, 0.48] &    0.07 & 873 \\
2013 & [-1.53, -0.76] & [-0.08, 0.48] &    0.06 & 851 \\
2014 & [-1.60, -0.82] & [-0.08, 0.49] &    0.07 & 852 \\
2015 & [-1.72, -0.91] & [-0.06, 0.52] &    0.07 & 852 \\
2016 & [-1.81, -0.99] & [-0.02, 0.58] &    0.08 & 865 \\
2017 & [-1.78, -0.96] & [-0.10, 0.51] &    0.08 & 865 \\
2018 & [-1.73, -0.87] & [-0.22, 0.43] &    0.08 & 866 \\
2019 & [-1.81, -0.95] & [-0.26, 0.40] &    0.09 & 874 \\
2020 & [-1.65, -0.84] & [-0.45, 0.21] &    0.11 & 874 \\
\end{tabular}
\end{table}

\begin{table}
\centering
\caption{Fits of calculated heterophobia corrections to GDP scaling deviations by year using the exposure index.}
\label{tab:expGDP}
\begin{tabular}{llll}

year &   $\pi$ 95\% CI & $R^{2}$ &   n \\
\hline \hline
2010 &  [-0.54, 0.26] & 0.00056 & 861 \\
2011 &  [-0.54, 0.32] & 0.00028 & 863 \\
2012 &  [-0.61, 0.26] & 0.00073 & 862 \\
2013 &  [-0.59, 0.27] & 0.00063 & 843 \\
2014 &  [-0.74, 0.14] & 0.00212 & 844 \\
2015 & [-0.88, -0.01] & 0.00471 & 847 \\
2016 & [-0.91, -0.05] & 0.00565 & 860 \\
2017 & [-0.88, -0.02] & 0.00487 & 859 \\
2018 &  [-0.87, 0.01] & 0.00434 & 858 \\
2019 & [-0.94, -0.10] & 0.00683 & 865 \\
2020 & [-1.01, -0.20] & 0.00992 & 868 \\
\end{tabular}
\end{table}

\begin{table}[bpht!]
\centering
\caption{Fits of calculated heterophobia adjustments to GDP scaling deviations by year with outliers included.}
\label{tab:outlierGDP}
\begin{tabular}{llll}

year &  $b^{het}$ 95\% CI & $R^{2}$ &   n \\
\hline \hline\\
2010 &  [-1.28, 0.04] & 0.00389 & 875 \\
2011 &  [-1.32, 0.07] & 0.00355 & 875 \\
2012 & [-1.50, -0.07] & 0.00528 & 875 \\
2013 & [-1.51, -0.09] & 0.00570 & 855 \\
2014 & [-1.72, -0.27] & 0.00834 & 855 \\
2015 & [-1.72, -0.37] & 0.01071 & 855 \\
2016 & [-1.65, -0.34] & 0.01011 & 868 \\
2017 & [-1.63, -0.28] & 0.00885 & 868 \\
2018 & [-1.67, -0.27] & 0.00847 & 868 \\
2019 & [-1.69, -0.36] & 0.01027 & 875 \\
2020 & [-1.60, -0.38] & 0.01147 & 875 \\

\end{tabular}
\end{table}

\begin{table}[bpht!]
\centering
\caption{Fits of calculated heterophobia adjustments to median income scaling deviations by year with outliers included.}
\label{tab:outlierInc}
\begin{tabular}{llll}

year & $b^{het}$ 95\% CI & $R^{2}$ &   n \\
\hline \hline\\
2010 & [-2.50, -1.19] &    0.06 & 885 \\
2011 & [-2.49, -1.16] &    0.06 & 885 \\
2012 & [-2.58, -1.26] &    0.06 & 885 \\
2013 & [-2.58, -1.27] &    0.07 & 864 \\
2014 & [-2.76, -1.44] &    0.07 & 864 \\
2015 & [-2.95, -1.61] &    0.08 & 864 \\
2016 & [-3.23, -1.90] &    0.09 & 877 \\
2017 & [-3.27, -1.94] &    0.10 & 877 \\
2018 & [-3.39, -2.06] &    0.10 & 876 \\
2019 & [-3.54, -2.20] &    0.11 & 885 \\
2020 & [-3.42, -2.12] &    0.11 & 885 \\

\end{tabular}
\end{table}
\end{document}